# Cluster-Head Rotation Approaches in Sensor Networks: A Review

*Rohit Pachlor*[*a] *Deepti Shrimankar*[b], *Kapil Kumar Nagwanshi*[*c], *Manish Paliwal*[c]

[a]Department of Computer Science and Engineering, GVP College of Engineering, Visakhapatnam AP 530048 INDIA
[b]Department of Computer Science and Engineering, Visvesvaraya National Institute of Technology, Nagpur, MH 440010, INDIA.
[c]Department of Information Technology, MPSTME, Narsee Monjee Institute of Management Studies, Shirpur, MH 425405, INDIA

ABSTRACT

Balanced energy consumption is a major research concern in sensor networks. If some sensor nodes spent energy rapidly compared to other sensor groups, then the energy consumption is not evenly distributed and the lifetime of the network is adversely affected. In cluster-based communication protocol, cluster-head nodes rapidly dissipate their energy compared to respective cluster members as it has additional duties of receiving data from cluster members and aggregation of received data. Clustering protocols periodically rotate the energy-intensive job of cluster-head among other sensors in order to evenly distribute the CH's energy load. In this paper, we review various cluster-head rotation approaches and classify them into different categories based on the number of participating nodes, rotation mechanism, and decision parameter(s). Finally, we discuss open research issues in developing an efficient cluster-head rotation approach, which reduces the premature death of cluster-head nodes and maximizes the network lifetime.

## 1. Introduction

It is expected that WSN will work independently for a long time with each sensor having a finite supply of energy. Recharging and replacing this finite source of sensor energy is almost impossible in many applications [1],[2]. For sensor nodes, it is therefore of utmost importance that they use their energy efficiently to prolong the lifetime of the network [3],[4],[5],[6]. The only three ways sensor nodes can reduce their energy consumption after the deployment of sensor networks are data aggregation, reduction of transmission power, and saving idle time energy.

- **Data aggregation** (see [7], [8], [9]): A sensor node sends and receives the number of packets directly proportional to the sensor's energy consumption. The data detected by nearby sensors is highly correlated in space. Data aggregation aggregates data from nearby sensors, eliminating redundant transmission of data. It reduces the network's use of bandwidth and energy consumption by eliminating redundant transmission.
- **Reducing transmission power** (see [9], [10] [11]): A sensor node sends and receives the number of packets directly proportional to the sensor's energy consumption. The data detected by nearby sensors is highly correlated in space. Data aggregation aggregates data from nearby sensors, eliminating redundant transmission of data. It reduces the network's use of bandwidth and energy consumption by eliminating redundant transmission.
- **Save idle time energy** (see [9], [10], [11], [12]): Duty cycling is used in WSN for two reasons: (1) sensors that periodically send their data can use duty cycles to save significant idle time energy. Sensor switches on its radio part at times depending on the availability of data. (2) Sensors can enter sleep mode and save their idle time if they do not need their presence to ensure coverage and connectivity.

All the above-mentioned goals can be achieved using a logical group-based topology. The literature presents the use of three cluster-based, chain-based, and tree-based, logical group-based topologies. The commonly used team topology is cluster-based, though, because it makes the network extremely scalable with less topology overheads [13], [14]. All network nodes are divided into several local clusters in the clustering protocol [15],[16]. -cluster comprises multiple nodes of cluster member (CM) and a node of cluster head (CH). The CH detects input from its surroundings in cluster-based WSN and collects information from nearby nodes as well. It then the aggregate the received information and communicate it to a BS. There are several rounds in the cluster-based protocol interaction system. Each communication round has two sub-phases: setup and transmission sub phase. The setup phase deals with clusters set up, cluster management. In data transmission phase, the sensor nodes transmit their data to the BS. This transmission can be single-hop or multi-hop. The CH works as proxy BS for CMs. In each communication round, the CMs send their data to the corresponding CH. The CH aggregates the received data and sends it to the BS. CH performs additional data collection and aggregation duties in a cluster compared to the CMs. According to related CMs[17], it therefore uses its energy quickly. The clustering protocols rotate CH's energy intensive work among other nodes in order to equally distribute the energy load of the CH in the network. The points chosen for CH task rotation have a major impact on the longevity of the network. If the rotation points of CH are not chosen correctly, premature death of CH nodes will increase. For the following reasons, an effective CH-rotation policy should minimize the premature death of a node in the form of CH:

- The premature death of CH ends the useful lifetime of all CMs of the respective cluster even if they have sufficient energy to do their work.
- If the premature death of CH occurs before the completion of communication round then it does not deliver any packets to the BS. In this case, the CH and CMs both waste their precious energy in an unsuccessful attempt to deliver the packet to the BS.
- In a cluster-based network, only the CH nodes communicate with the BS. In this case, if a node prematurely dies as a cluster-head, then it directly affects the total number of packets received at the BS.

*1.1. Our Contributions*

The scope of this review paper is limited to the "balanced energy consumption" aspect of clustering, not the clustering itself. Many review and survey articles are already published on cluster-based communication in WSN [15], [16], [17], [18], [19], [20], [21], [22], [23], [24]. The "balanced energy consumption" is a key aspect of clustering, which has a huge impact on the overall network lifetime. Therefore, we choose the balance energy consumption as a review topic for this research. In this paper, we review various approaches for rotating the CH and classify them into various categories based on the number of participating nodes, rotation mechanism, and decision parameter(s). We discuss the benefits of one approach on another approach(s). And finally, we discuss open research issues in developing an efficient CH-rotation approach, which reduces the premature death of cluster-head nodes and maximizes the network lifetime. This is the first ever survey paper which reviews the CH rotation policies proposed for sensor networks. Many IoT applications also use clustering as an energy conservation tool [25], [26], [27], [28]. For example, in public safety networks (PSN) or public safety communications, emergency first responders (EFRs) such as rescue officers, firefighters, medical officers and police officers use a variety of wireless communication technologies to timely coordinate situational awareness among themselves and a control center (CC). The communication network of PSNs includes battery-operated nodes that communicate directly with the CC. This single hop communication between nodes with CC makes the network energy-inefficient due to the large distances between the nodes and the CC. In this situation, clustering can be used for better power management. Therefore, the proposed review paper is not only helpful in WSN but also for IoT applications which require efficient power management of wireless nodes.

*1.2. Paper outline*

The rest of the paper is organized as follows: Section II presents and classifies CH rotation approaches. Section III describes the challenges of CH-rotation policies. Finally, section VI concludes the findings.

## 2. Rotation of Cluster Heads

*2.1. Classification based on Rotation Mechanism*

The cluster-head rotation can be defined as centralized and decentralized CH rotation. If the base station coordinates cluster-head rotation, then the cluster-head rotation is centralized otherwise it is decentralized. The cost of the centralized cluster-head rotation is higher than the decentralized cluster-head rotation. LEACH is the first clustering protocol proposed for WSN [1], [29], [30]. In LEACH, after every communication round the network is re-clustered by randomly selecting the new cluster-head nodes. In LEACH-C and LEACH-F, the base station selects the new cluster-head nodes after every communication round. The nodes with highest residual energy are selected as new cluster-heads [1], [30]..

*2.2. Classification based on Decision Parameter(s)*

The base station/cluster-head rotates the job of cluster-heads among other network nodes primarily based on pre-defined criteria(s). These pre-defined criteria(s) place a condition on some network parameter(s). For example, if the remaining energy of cluster-head goes under a pre-set threshold. If at any time the parameter completes the specified condition, then the base station / cluster-head rotates cluster-head job. Based on the decision parameter(s), the cluster-head rotation technique(s) can be categorized as EDCR, TDCR, and HETDCR.

**Energy-driven CH-rotation (EDCR)** (see [31], [32]): In EDCR, the cluster-head rotation decision is made on the basis of an energy parameter condition. If the residual energy of current cluster-head goes under a preset threshold or if the average cluster member energy goes under a preset threshold or if the residual energy of cluster-head goes under the average cluster member energy, then the base station/cluster-head rotates the job of CHs. EDCR can be centralized or decentralized. In case of centralized cluster-head rotation, if the cluster-head's remaining energy goes below the threshold (or any preset condition on energy parameter gets satisfied), then it triggers the cluster-head rotation event by informing the base station. The current cluster-head informs the base station that it is not able to perform its responsibilities as a cluster-head any longer. Later on, the base station informs this to all other cluster-heads of the network and starts the cluster-head rotation. In DCR, if any condition on energy parameter gets satisfied, the CH itself takes the decision of CH-rotation.

For cluster-head rotation, the energy-driven CH rotation uses or dynamic threshold. The cluster-head rotation based on static threshold increases the premature death of cluster-head nodes due to fact that it does not update itself according to the current load of cluster-head. On the other hand, dynamic threshold is responsive to the current load of the cluster-head and update itself according to the current load of CH. In addition, rotation in EDCR becomes increasingly frequent as the average residual energy of nodes reduces. In [33], the author(s) have shown via simulation that the energy efficiency of energy driven CH rotation is very high when the remaining energy of nodes is abundant and it decreases as the remaining energy of nodes decreases

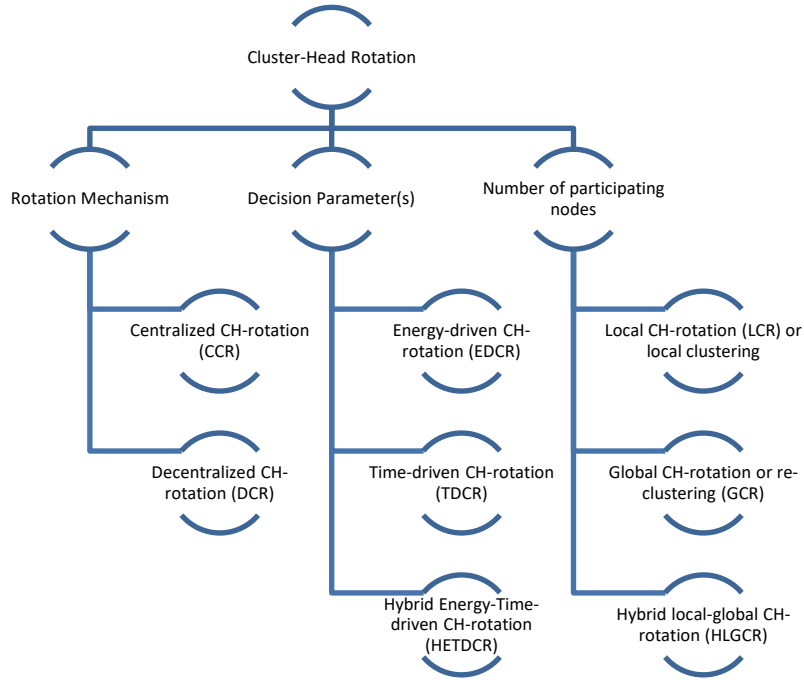

**Fig. 1 - Taxonomy of approaches used for rotating the cluster heads position.**

**Time-driven CH-rotation (TDCR)** (see [1], [29], [30]): In time-driven CH rotation, the cluster-head rotation is periodic. The cluster-head rotation decision is based on time parameter i.e. number of communication rounds. The CH-rotation can occur after every communication round or after multiple communication rounds. If a cluster-head divides its residual energy by energy needed to successfully complete one round, then it can estimate the number of rounds that it can successfully completed before it dies. If we rotate cluster-head just after the projected number of rounds, then the cluster-head will die before rotation occurs. Therefore, the job of CHs can be rotated by subtracting one or several rounds from the estimated number of rounds.

The time-driven rotation balances the energy consumption of nodes by rotating the load of CH evenly among other nodes of the network. Unlike EDCR, TDCR has low energy efficiency when the residual energy of nodes is high because it does not consider the residual energy of CH in rotation decision. The TDCR is not suitable for a heterogeneous network, where the initial energy of all nodes is not identical. In [33], author(s) have shown via simulation that TDCR maintains the constant energy-efficiency regardless of network lifetime.

**Table 1 - Comparison of Cluster-head Rotation approaches.**

| Comparison | Type of CH-rotation (CR) | | | | | |
|---|---|---|---|---|---|---|
| | **CCR** | **DCR** | **EDCR** | **TDCR** | **LCR** | **GCR** |
| CH-rotation Process | BS coordinates the CH-rotation | CH coordinates the CH-rotation | An energy-driven threshold is set on the residual energy of CH or average cluster member energy. | The CH-rotation is periodic and it is performed after the predetermined number of communication round(s). | The role of CH is rotated locally among the associate CMs. | The role of CH is rotated locally among the all the nodes of the network. |
| Pros | BS ensures high synchronization among clusters. | Rotation overhead is low. | The more the residual energy of the CH has, the higher the energy efficiency can be obtained. | Constant energy efficiency can be maintained regardless of network lifetime. | Rotation cost is very low as less number of nodes participates in the rotation. | Energy consumption is a balance among all the nodes of the network. Clusters are dynamic |
| Cons | Rotation overhead is high because overhead to reach BS gets added and it may become very high because as the distance to BS increases. | Difficult to keep synchronization among clusters. | 1. Low energy efficiency if the residual energy of CHs is inadequate. 2. Static threshold lead to premature death of CHs. | Low energy efficiency when the residual energy of CH is high because it does not consider the residual energy of CH in rotation decision. | 1. Clusters are static 2. Energy consumption of cluster members is not balanced | Rotation cost is very low as all the nodes of the network participates in the rotation. |

**Hybrid Energy-Time-driven CH-rotation (HETDCR)** (see [33]) : Depending on the preset condition, hybrid energy-time-driven CH-rotation switches the rotation strategy from energy-driven to time-driven or vice-versa. It combines the benefits of both energy-driven and time-driven cluster-head rotation. In [33], the author(s) have shown that the EDCR's energy efficiency is low than that of TDCR when the remaining energy of nodes is very low and it is very high compared to TDCR when the nodes have abundant remaining energy. They have analytically set an energy limit on the average remaining energy of nodes. EDCR outperforms TDCR when the average remaining energy of nodes exceeds the threshold and TDCR outperforms EDCR when it goes under the threshold. Based on this threshold, they have proposed a HETDCR technique, which switches the rotation technique from EDCR to TDCR as the remaining energy of nodes goes under the threshold.

*2.3. Classification based on the number of participating nodes*

Based on the number of participating nodes the CH-rotation approaches can be classified as LCR also called local clustering, GCR also called re-clustering and HLGCR.

**Global CH-rotation or re-clustering (GCR)** [1], [30]: In re-clustering, all nodes of the network participate in creating new clusters using a preset cluster-head selection and cluster membership criteria. As soon as the cluster-heads are selected using the preset cluster-head selection criteria, they broadcast their cluster-head status across their communication range. The non-cluster-head nodes join one of the cluster head nodes based on the received signal strength. After every non cluster head node joins an optimal cluster head for communicating its sensed data, the cluster-head node creates and broadcast a TDMA schedule into its cluster for data communication. Each member transmits its data to its associated cluster-head according to the intended slot specified by the TDMA schedule.

The cluster-head selection, the cluster setup, and the formation and propagation of TDMA schedule are an overhead over the real data communication. And since all nodes of the network participate in re-clustering, the number of message exchanges is higher in the setup phase and hence the cost of re-clustering and performing clustering in each communication round only increases this overhead. Re-clustering can be centralized or decentralized or energy-driven or time driven.

**Local CH-rotation (LCR) or local clustering** (see (LC) [29], [31], [32]): local CH-rotation reduces the cost of re-clustering. In local CH-rotation, only the cluster members of particular cluster take part in selection of new cluster head for the next round. Usually, if the remaining energy of current cluster head goes under the preset threshold, it triggers cluster-head rotation. In cluster-head rotation, usually, the current cluster-head selects the new cluster head for next communication round depending on some preset cluster-head selection criteria such as remaining energy of cluster-head, inter-cluster distance, node-degree etc. After the selection of new cluster-head, the current cluster-heads becomes a normal member of the cluster.

In CH-rotation, the cluster boundaries are static and CMs play the role of CH, one by one. The overhead cost of CH-rotation is less than the re-clustering overhead but it creates the static clusters. Therefore, it is not practical for any type of dynamic system because it does not allow new nodes to be added to the system and does not adjust its behavior for dead nodes. LEACH-F is the first local-CH rotation based clustering protocol for sensor networks but the cluster-head rotation is centralized [29].

**Hybrid local-global CH-rotation (HLGCR)** (see [32]): The local CH-rotation reduces the rotation overhead by rotating the cluster-head jobs locally among the associated cluster-members and the re-clustering redefines the cluster boundaries by creating the new clusters periodically. The Hybrid local-global CH-rotation combines the benefits of both local CH-rotation & global CH-rotation. The hybrid technique uses preset criteria for rotating the job of cluster-head. If the remaining energy of current cluster-heads goes under the preset rotation threshold, then the current cluster head triggers the rotation of cluster-head and selects one of the members of cluster as a new cluster-head for the next round. If the current cluster node does not find any suitable member node that can take the responsibility of cluster-head, then it calls for re-clustering. In [32], the author(s) rotates the role of cluster-head among associated CMs until the average cluster-member energy does not fall below a preset threshold. If the ACME falls below the threshold, the re-clustering is triggered.

Table 1. compares the CH-rotation approaches and lists the pros and cons of each rotation approach. Table 2 shows the comparison of prominent clustering protocols proposed for WSN since 2000 with respect to the type of CH-rotation approach used. LEACH is proposed in the year 2000, the first clustering protocol for WSN. The main aim of Table 2 is to classify, the rotation approaches used in literature since 2000, as CCR, DCR, EDCR, TDCR, HETCR, LCR, GCR, HLGCR according to the taxonomy shown in figure 1.

## 3. Challenges with CH rotation approaches

In this section, we discuss the problems of existing CH-rotation protocols.

**The problem of static clusters:** The clusters remain constant in LCR throughout the lifespan of the network. It is therefore not suitable for a dynamic system based on newly added and dead nodes that changes its behavior. Consequently, GCR must be used to rotate CHs.

**High energy dissipation of CMs:** LCR can suffer from high-energy non-CH node (CM) dissipation. If a new CH is chosen solely on the basis of the residual energy parameters and the range within the cluster is not considered for selection, the newly picked CH may be far from other CMs.

***The problem of the static threshold for CH-rotation:*** In LCR, the cluster boundaries are static and CMs play the role of CH, one by one. Usually, when the residual energy of CH falls below a threshold, it triggers CH-rotation. The energy threshold for rotating CH can be static or dynamic. The static threshold does not consider the current energy load of CH. Thus, it increases the premature death of the CH nodes. Let consider a cluster of *m* nodes with one of the nodes acting as CH and *(m-1)* nodes acting as CM. $E_{R\_CCH}$ is the residual energy of CCH, $E_{Th\_CR}$ is the static energy threshold for CH-rotation.

$E_{L1R}$ is the total amount of energy load required by CCH to successfully complete one communication round. The equation (1) represents the CH-rotation operation.

$$f(E_{R\_CCH}, E_{Th\_CR}, E_{L1R}) = \begin{cases} CCH \rightarrow CH & if \quad E_{Th\_CR} < E_{R\_CCH} > E_{L1R} (I) \\ CCH \rightarrow CH & if \quad E_{Th\_CR} < E_{R\_CCH} < E_{L1R} (II) \\ CCH \rightarrow CH-rotation & if \quad E_{R\_CCH} \leq E_{Th\_CR} (III) \end{cases} \quad (1)$$

In (I) case of (1), CCH's residual energy is more than the CH-rotation static energy threshold and thus the CCH can continue its CH job for the next round of communication. Additionally, CCH's remaining energy is more than the energy needed by CCH to complete one interaction round successfully. Therefore, if the CCH completes the work in CH for the next round of contact, it will not die prematurely. In (II) case of (4), CCH's residual energy is more than the CH-rotation fixed energy limit but less than the energy needed to complete one interaction round successfully. Therefore, if the CCH continues its job as CH for the next communication round, it will die prematurely. This is mostly because the CH-rotation limit is constant. If it were fluid, it would vary depending on the actual CH load, so CH's death is not premature.

*The problem of an overloaded cluster:* Each member of LCR assumes CH's responsibility. Each participant completes k rounds as a CH before CH-rotation just for simplicity's sake. Therefore, due to the low-energy level, no cluster member will be able to take responsibility for CH after km m rounds. These clusters are called clusters that are crowded or bloated. If we choose the largest residual energy member or any member as CH in an overloaded cluster, it will die prematurely.

*Re-clustering frequency and overhead:* GCR allows to redefine the boundaries of the cluster using re-clustering and enables to choose the new CHs for each cluster. But it suffers from two major problems: (1) performing clustering in each communication round increases the setup overhead. (2) In re-clustering, all nodes of the network participate in forming new clusters and selecting new CHs for the network. This increases the number of control packets exchanged in the network. In addition, if the clustering approach is centralized then the overhead of re-clustering increases more (worst, when the BS located far away from the sensing field). The overhead of centralized clustering is higher than distributed clustering as the additional overhead to reach BS gets added to the total overhead of clustering.

## 4. Discussions

Can a local clustering policy be designed that completely eliminates the need for re-clustering but retains all the benefits of re-clustering? In addition, the issue of high energy dissipation of CMs, fixed threshold and crowded cluster should also be discussed. Authors also proposed a load-adaptive solution to CH rotation in[ 63]. Every CH calculates the number of contact rounds in this method that it can successfully complete before dying based on the existing CH job load. A local cluster adjustment policy has been proposed in[ 64] by authors that eliminates the need for re-clustering. It also takes care of high-energy CMs dissipation, static threshold and cluster overload. The rule of cluster modification makes a CM after CH-rotation to join another cluster. If the dissipation of energy to enter the cluster's CH is less than the dissipation of energy to hit the current cluster's CH. It allows cluster borders to be redefined and prevents re-clustering. In[ 65], the authors proposed a local cluster splitting policy dividing a cluster into two sub-clusters if no CH member can assume CH's responsibility. It also redefines cluster boundaries by dividing the clusters into two and thus eliminates re-clustering.

## 5. Conclusions

In a cluster-based sensor network, CH rotation achieves balanced node energy consumption. In this paper, we reviewed and classified the various existing approaches to rotating CH's load as centralized, decentralized, energy-driven, time-driven, hybrid, time-driven, local, global, and hybrid local-global CH-rotation policies. In various clustering protocols proposed for WSNs, we have summarized and classified techniques used to rotate CH's role. In developing an effective CH-rotation policy, we have outlined the most important problems and addressed some recent approaches for it. An active CH-rotation policy would reduce the overhead set-up to rotate CH's position. It should also address the issue of high-energy dissipation of CMs, fixed thresholds and overloaded clusters.